\documentclass{llncs}
\usepackage[final]{pdfpages} 
\usepackage{wrapfig}
\usepackage{color}
\usepackage{tikz}
\usetikzlibrary{arrows,automata,shapes,backgrounds,fit,decorations.pathreplacing,patterns}

\usepackage{pgfplots}

\usepackage{amssymb,amsmath}
\usepackage{stmaryrd,wasysym}
\usepackage{xspace}
\usepackage{url}
\usepackage{extarrows}
\usepackage{mathtools}
\usepackage{paralist}
\usepackage{listings}
\usepackage{subfigure}

\makeatletter
   \def\marginparright{\@mparswitchfalse}
   \def\marginparoutside{\@mparswitchtrue}
\makeatother
\def\comment#1{}

\def\withcomments{
  \setlength{\marginparwidth}{2.5in}
  \addtolength{\oddsidemargin}{-1in}
  \addtolength{\evensidemargin}{-1in}
  \newcounter{mycommentcounter}
  \marginparright
  \def\comment##1{\refstepcounter{mycommentcounter}%
    \ifhmode%
    \unskip%
    {\dimen1=\baselineskip \divide\dimen1 by 2 %
      \raise\dimen1\llap{\tiny {\bf \color{red}*-\themycommentcounter-*}}}\fi%
    \marginpar{\renewcommand{\baselinestretch}{0.8}%
      \footnotesize [\themycommentcounter]: \raggedright ##1}}
  \date{\framebox{Draft of \today}}
}

\renewenvironment{proof}[1][Proof]
{\begin{trivlist} \item \textbf{#1}\ \ }{\qed\end{trivlist} }

\def\Q{\ensuremath{\mathcal{Q}}}

\newcommand{\set}[1]{\ensuremath{\{ #1 \}}}
\newcommand{\true}{1}
\newcommand{\false}{0}
\newcommand{\match}{\ensuremath{\true_{\rm T}}\xspace}
\newcommand{\nomatch}{\ensuremath{\false_{\rm T}}\xspace}
\newcommand{\noreqmatch}{\ensuremath{\bot_{\rm T}}\xspace}
\newcommand{\allow}{\ensuremath{\true_{\rm P}}\xspace}
\newcommand{\deny}{\ensuremath{\false_{\rm P}}\xspace}
\newcommand{\na}{\ensuremath{\bot_{\rm P}}\xspace}

\newcommand{\tand}{\mathbin{\mathsf{and}_{\rm T}}}

\newcommand{\tnot}{\mathop{\mathsf{not}_{\rm T}}}

\newcommand{\pand}{\mathbin{\mathsf{and}_{\rm P}}}

\newcommand{\pnot}{\mathop{\mathsf{not}_{\rm P}}}
\newcommand{\dbd}{\mathop{\mathsf{dbd}_{\rm P}}}

\newcommand{\tweakening}{\mathop{\mathsf{opt}_{\rm T}}}
\def\topt{\tweakening}
\newcommand{\kweakening}{\mathop{\sim}}
\newcommand{\sem}[1]{\ensuremath{\llbracket #1 \rrbracket}}
\newcommand{\tsem}[1]{\sem{#1}_{\rm T}}
\newcommand{\psem}[1]{\sem{#1}_{\rm P}}

\newcommand{\weakkland}{\sqcap}
\newcommand{\weakklor}{\sqcup}
\newcommand{\strongkland}{\mathbin{\tilde{\sqcap}}}
\newcommand{\strongklor}{\mathbin{\tilde{\sqcup}}}

\newcommand{\Att}{\ensuremath{\mathcal{A}}}
\newcommand{\fv}{\ensuremath{\mathit{fv}}}
\newcommand{\nf}{\ensuremath{\mathsf{nf}}}
\newcommand{\NF}{\ensuremath{\mathsf{NF}}}

\newcommand{\nat}{\ensuremath{\mathbf{nat}}\xspace}
\newcommand{\fr}{\ensuremath{\mathsf{FR}}\xspace}
\newcommand{\at}{\ensuremath{\mathsf{AT}}\xspace}
\newcommand{\de}{\ensuremath{\mathsf{DE}}\xspace}
\newcommand{\uk}{\ensuremath{\mathsf{UK}}\xspace}
\newcommand{\nv}{\ensuremath{\mathsf{NV}}\xspace}

\newcommand{\ptackle}{PTaCL\xspace}
\newcommand{\ATRAP}{ATRAP\xspace}
\lstdefinelanguage{isa} {alsoletter={\#, \&},
  mathescape=true,
  basicstyle=\small,
  escapechar=\@,
  boxpos=c,
  morekeywords= {theorem, lemma, apply, by, constdefs, definition, where,
    infixl, types, consts, primrec, have, from, show, let,  proof, qed, is,
    sorry, with, assume, fix, thus, hence, datatype, if, then, else, in, case,
    theory, of, imports, begin
  },
  emph={[2]int, char, string, self, boolean},
  emphstyle={[2]\it},
literate=
  {->}{{$\rightarrow$}}2
  {=>}{{$\Rightarrow$}}2
  {-->}{{$\longrightarrow$}}2
  {\\forall}{{$\forall$}}2
  {\\exist}{{$\exists$}}2
  {\\exists}{{$\exists$}}2
  {AND}{{$\& \&$}}3
  {ALL}{{$\forall$}}2
  {EX}{{$\exists$}}2
  {\%}{{$\lambda$}}1
  {\\\/}{{$\sqcap$}}2
  {|-}{{$\vdash$}}2
  {==}{{$\equiv$}}2
  {\\<Longrightarrow>}{{$\Rightarrow$}}2
  {\\<longrightarrow>}{{$\rightarrow$}}2
  {\\<Rightarrow>}{{$\Rightarrow$}}2
  {\\<forall>}{{$\forall$}}2
  {\\<exists>}{{$\exists$}}2
  {\\<equiv>}{{$\equiv$}}2
  {\\<noteq>}{{$\neq$}}2
  {\\<not>}{{$\lnot$}}2
  {\\<in>}{{$\in$}}2
  {\\<or>}{{$\lor$}}2
  {\\<subseteq>}{{$\subseteq$}}2
  {\\<preceq>}{{$\preceq$}}2
  {==>}{{$\Rightarrow$}}2
  {~}{{$\neg$}}1
  {~=}{{$\neq$}}2
  {:=}{{\tt :=}}2
  {;}{{\tt ;}}1
  { graph }{{ $G$ }}2
}

\lstnewenvironment{isaenv} {
  \lstset{extendedchars=true,columns=flexible}
  \lstset{language=isa,deletestring=[b]', basicstyle={\footnotesize \sf}}%
} {}

\lstdefinelanguage{maude} {alsoletter={\#, \&},
  mathescape=true,
  basicstyle=\small,
  escapechar=\@,
  boxpos=c,
  morekeywords= {theorem, lemma, apply, by, constdefs, definition, where,
    infixl, types, consts, primrec, have, from, show, let,  proof, qed, is,
    sorry, with, assume, fix, thus, hence, datatype, if, then, else, in, case,
    of
  },
  emph={[2]int, char, string, self, boolean},
  emphstyle={[2]\it},
literate=
  {~>}{{$\sim>$}}2
}

\lstnewenvironment{maudeenv} {
  \lstset{extendedchars=true,columns=flexible}
  \lstset{language=maude,deletestring=[b]', basicstyle={\footnotesize \sf}}%
} {}

\lstnewenvironment{atrapenv} {
  \lstset{extendedchars=true,columns=flexible}
  \lstset{language=java,deletestring=[b]', basicstyle={\footnotesize \tt}}%
} {}

\newtheorem{prop}{Proposition}
\newtheorem{Thm}{Theorem}

\newcommand{\isa}[1]{{\lstinline[language=isa, basicstyle={\footnotesize \sf} ]@#1@}}
\newcommand{\maude}[1]{{\lstinline[language=maude, basicstyle={\footnotesize \sf} ]+#1+}}
\newcommand{\java}[1]{{\lstinline[language=java, basicstyle={\footnotesize \sf} ]@#1@}}
\newcommand{\atrap}[1]{{\lstinline[mathescape=true,language=java, basicstyle={\footnotesize \sf} ]@#1@}}

\begin{document}
\mainmatter  

\title{Automated Certification of Authorisation Policy
  Resistance\thanks{Work partially supported by the International
    Exchange Scheme of the Royal Society and conducted in part at
    Imperial College London where A. Griesmayer was supported by the
    Marie Curie Fellowship ``DiVerMAS'' (FP7-PEOPLE-252184). }}

    \titlerunning{Certification of Policy Resistance}

    %
    %
    
    \author{Andreas Griesmayer\inst{1} \and Charles Morisset\inst{2}
      }
\institute{
  ARM, Cambridge, UK\\
  \email{andreas.griesmayer@arm.com}
 \and
 Center for Cybercrime and Computer Security\\
 School of Computing Science, Newcastle University, UK \\
 \email{charles.morisset@ncl.ac.uk}
}

    \authorrunning{Certification of Policy Resistance}

 
    \maketitle

\begin{abstract}
  Attribute-based Access Control (ABAC) extends traditional Access Control by considering an access request as a set of pairs \textit{attribute name}-\textit{value}, making it particularly useful in the context of open and distributed systems, where security relevant information can be collected from different sources. However, ABAC enables {\em attribute hiding attacks}, allowing an attacker to gain some access by withholding information. 

  In this paper, we first introduce the notion of policy {\em resistance} to  attribute hiding attacks. We then propose the tool \ATRAP (Automatic Term Rewriting for Authorisation Policies), based on the recent formal ABAC language \ptackle, which first automatically searches for resistance counter-examples using Maude, and then automatically searches for an Isabelle proof of resistance. We illustrate our approach with two simple examples of policies and propose an evaluation of \ATRAP performances. 
\end{abstract}

\section{Introduction}
\label{sec:introduction}

An authorisation policy for a security mechanism is a document describing which user requests are authorised and which ones are denied. 
Many languages exist in the literature to define policies, most of them considering a request as a triple subject-object-mode. 
Recent approaches~\cite{Rao:2009:AFI:1542207.1542218,XACML3,post2012} consider more expressive requests, consisting of a set of pairs of attribute name and attribute values, thus defining the model known as Attribute Based Access Control (ABAC). 

A major feature of ABAC is its ability to gather attribute information from different sources.  This is essential in open and distributed systems, which indeed often lack a central security point providing all required information.  In order to make a security decision, such systems need to combine information from different sources including the user herself (e.g., personal identification), the environment (e.g., date or location), the data requested (e.g., relationship with other objects) and  other users (e.g., availability of higher-ranked users). 

Delegating the retrieval of security relevant information clearly brings more flexibility in open systems, but also raises the problem of information withholding, malicious or not. Tschantz and Krishnamurthi introduce in~\cite{confsacmatTschantzK06} the notion of safety, such that a policy is safe when ``incomplete requests should only result in a grant of access if the complete one would have''. More recently, Crampton and Morisset identify in~\cite{post2012} the notion of an {\em attribute hiding attack}, where a user 
hides some attribute values in order to get access to a resource she would be denied otherwise. 

As a running example to illustrate our research problem, consider a fictional organisation, sponsored by both Austria and France, where documents can be submitted and reviewed.  The organisation has a simple conflict-of-interest policy stating that a document cannot be reviewed by someone with the same nationality as the submitter. The implementation of such a policy might lead to attribute hiding attacks: 
Consider a policy for a document submitted by an Austrian member that states that it cannot be accessed by another Austrian member; If an Austrian attacker is able to hide her nationality, for instance by corrupting the corresponding data, then she could access the document. Such an attack is particularly relevant when a user can have multiple nationalities.  

In this paper, we focus on the behaviour of authorisation policies when facing attribute hiding attacks. More precisely, we introduce the notion of {\em resistance}: a policy is resistant when every query obtained by adding information to an allowed query is also allowed. This definition generates the following research questions: 
\begin{enumerate}
\item Is it possible to automatically detect whether a policy is resistant or not? 
\item If a policy is not resistant, can we exhibit a counter-example? 
\item If a policy is resistant, can we construct a formal proof of it? 
\end{enumerate}
The main contribution of this paper is to positively answer these
questions, at least in a partial way. In order to do so, we present
the tool \ATRAP (Automated Term Rewriting for Authorisation Policies),
which combines the term-rewriting tool Maude~\cite{clavel2007all} and the proof assistant
Isabelle/Isar~\cite{DBLP:books/sp/NipkowPW02,Wenzel07theisabelle/isar} to analyse the resistance of PTaCL policies.  The core of the
tool is written in Java and handles the communication between Maude
and Isabelle by generating the respective inputs and interpreting
their results.  \ATRAP is capable of generating counter-examples for
non resistant policies, and of building an Isabelle proof for
resistant policies. Although \ATRAP is sound, i.e., all
counter-examples and proofs generated are indeed correct, \ATRAP is not
(yet) complete, as it may fail in some cases to find a counter-example
in a reasonable amount of time or to build a proof.

The rest of the paper is structured as follows: in Section~\ref{sec:ptacl}. we present the language \ptackle and introduce its evaluation in Maude. In Section~\ref{sec:resistance}, we define the notion of policy resistance and show how this property can be verified in practice. In Section~\ref{sec:counterexample}, we describe our approach to find a counter-example of resistance, and  in Section~\ref{sec:certification}, we present how a {\em proof} of resistance can be automatically generated, after which we conclude
and present future work in Section~\ref{sec:conclusion}. 

\subsection*{Related work}

The notion of resistance introduced in this paper is related to that of safety given in~\cite{confsacmatTschantzK06}, which roughly states that the evaluation of a request should be ``lower'' than that of a request with strictly more information. This notion is close to that of (weak)-monotonicity~\cite{post2012}.  As shown in Section~\ref{sec:resistance}, a (weakly) monotonic policy is also resistant, while the converse does not always hold. 

There exist several approaches using model-checking to analyse access control policies~\cite{DBLP:conf/isw/GuelevRS04,Zhang:2008:SVA:1370684.1370685}.  For instance, 
SMT solving can be used to check whether a given request can eventually be granted with a particular role in Administrative-RBAC policies~\cite{Alberti:2011:ESA:1966913.1966935,DBLP:journals/jcs/ArmandoR12}. Similarly, 
XACML policies can be automatically compared using a SAT solver~\cite{Hughes:2008:AVA:1459278.1459282}. 
To the best of our knowledge, we are the first to automatically analyse the resistance of policy, and also to allow for the generation of a structured proof. 

The automation of \ATRAP relies on term-rewriting, which has already been considered for the formalisation of access control~\cite{Barker:2006:TRA:2090485.2090498,Dougherty:2007:MAC:2393847.2393899}, leading to the analysis of rewrite-based policies~\cite{Kirchner:2009:ARA:1519544.1519756}. In particular, Bertolissi and Uttha propose in~\cite{Bertolissi:2013:AAR:2428116.2428125} to relate the properties of a rewrite system, such as totality or consistency, with those of an access control policy encoded in this system. This approach allow them to use the rewrite system CiME to generate proof certificates for these properties in the proof assistant CoQ. We follow a similar objective here, in that we generate proof certificates using a rewrite system, however we focus on the notion of resistance, which is not a property of the rewrite system itself, but of the policy. 

\ATRAP relies on the encoding of \ptackle in Isabelle, mentioned in~\cite{post2012}, in order to generate a proof of resistance. Brucker et al. present in~\cite{Brucker:2011:AMT:1998441.1998461} an encoding of an access control model into HOL, in the context of healthcare policies. Finally, it is worth mentioning that some of the techniques used in \ATRAP are inspired by previous work~\cite{Griesmayer:2013:FAC:2443461.2443470}, where the basic idea of using term-rewriting to generate proofs is used in the context of program refinement.

\section{\ptackle}
\label{sec:ptacl}
XACML 3.0~\cite{XACML3} is an OASIS standard for representing authorisation policies, and given an access request, its complete request evaluation cycle can be summarised as follows: 
\begin{inparaenum}[(i)]
\item the request is submitted to the Policy Enforcement Point (PEP);
\item the PEP forwards the request to the Context Handler (CH); 
\item the CH collects all attributes necessary to the evaluation of the request; 
\item the CH forwards the complete request to the Policy Decision Point (PDP); 
\item the PDP evaluates the complete request and returns the corresponding decision to the CH, which returns it to the PEP. 
\end{inparaenum}

\ptackle~\cite{post2012} formalises the evaluation of the request by the PDP, which considers each request as complete, or more precisely, cannot make a distinction between incomplete and complete requests. In other words, if the CH is unable to collect some attributes, and forwards an incomplete request to the PDP, this request is evaluated in the same way than a complete one. 

We present in this section the language PTaCL through the description of an illustrative example. We introduce the different definitions required for the understanding of this example, and refer to~\cite{post2012} for further details about the language. We take as example the one given in the introduction, where the access to a document is based on the nationality of the requester.

\subsection{3-valued Logic}

The 3-valued logic extends the traditional Boolean logic $\set{1, 0}$, where $1$ represents $\mathit{true}$
and $0$ represents $\mathit{false}$, by considering an additional value $\bot$~\cite{kleene:intr50}. 
The usual Boolean operators, such as the conjunction, disjunction, negation, etc, can be extended to the set 
$\set{\true,\false,\bot}$, as shown in Fig.~\ref{fig:applicability-logical-operators}. The weak operators consider the value $\bot$ as absorbing, while the strong ones ``resolve'' $\bot$ as much as possible. 

The logic $\set{\true, \false, \bot, \strongklor, \kweakening, \lnot}$ is proven in~\cite{post2012} to be functionally complete, using a result from Jobe~\cite{Jobe62}, which means that any logical operator can be built from these operators and constants. 
In the following, we sometimes use the set $\set{\true, \false, \bot, \strongkland, \kweakening, \lnot}$, since 
$x \strongkland y = \lnot (\lnot x \strongklor \lnot y)$. In addition, we use the three valued logic both to represent the result of target evaluation  and the result of policy composition. In order to avoid any confusion, we use $\set{\match, \nomatch, \noreqmatch}$ for the former, and $\set{\allow, \deny, \na}$ for the latter, whose meaning will be given in due course. 

\begin{figure*}[tp]
\caption{Binary and unary operators on the target decision set $\set{\true,\false,\bot}$}\label{fig:applicability-logical-operators}
  \subfigure[Weak operators]{
    \begin{minipage}{.185\textwidth}\small
      \[
        \begin{array}{c|ccc}
          \weakkland & \true & \false & \bot \\
        \hline
          \true & \true & \false & \bot \\
          \false & \false & \false & \bot \\
          \bot & \bot & \bot & \bot \\
        \end{array}
      \]
    \end{minipage}
    \begin{minipage}{.185\textwidth}\small
      \[
        \begin{array}{c|ccc}
          \weakklor & \true & \false & \bot \\
        \hline
          \true & \true & \true & \bot \\
          \false & \true & \false & \bot \\
          \bot & \bot & \bot & \bot \\
        \end{array}
      \]
    \end{minipage}}
  \subfigure[Strong operators]{
    \begin{minipage}{.185\textwidth}
      \[
        \begin{array}{c|ccc}\small
          \strongkland & \true & \false & \bot \\
        \hline
          \true & \true & \false & \bot \\
          \false & \false & \false & \false \\
          \bot & \bot & \false & \bot \\
        \end{array}
      \]
    \end{minipage}
    \begin{minipage}{.185\textwidth}\small
      \[
        \begin{array}{c|ccc}
          \strongklor & \true & \false & \bot \\
        \hline
          \true & \true & \true & \true \\
          \false & \true & \false & \bot \\
          \bot & \true & \bot & \bot \\
        \end{array}
      \]
    \end{minipage}}
    \subfigure[Unary]{
      \begin{minipage}{.185\textwidth}\small
        \[
          \begin{array}{c|c|c}
            X & \lnot X & \kweakening X \\ \hline
            \true & \false & \true \\
            \false & \true & \false \\
            \bot & \bot & \false \\
          \end{array}
        \]
      \end{minipage}}
\end{figure*}

\subsection{Target and Policy}
\label{sec:ptackle}

Following recent work~\cite{Rao:2009:AFI:1542207.1542218,XACML3}, \ptackle is attribute-based, meaning that a request is modeled as a set of attribute name-value pairs. Our running example uses an attribute \nat, whose value can be either \fr or \at. For instance, the request $\set{(\nat, \fr)}$ represents a request made by a French national. 

In addition, \ptackle is target-based~\cite{bona:alge02,DBLP:journals/tissec/BrunsH11,CH10,xacml2.0,XACML3,wije:prop03}, 
meaning that an access control policy contains a target that specifies the requests to which the policy is applicable, and a body (either a single decision or another policy) describing how applicable requests should be evaluated. 

In its simplest form, an {\em atomic target} is a pair $(n, v)$, where $n$ is an attribute name and $v$ is an attribute value. For instance, the target $(\nat, \fr)$ evaluates to $\match$ (match) if the request contains $(\nat, \fr)$, to $\nomatch$ (no-match) if it contains  $(\nat, \at)$, but not $(\nat, \fr)$, and to 
$\noreqmatch$ (indeterminate) if it does not contain any value for the attribute \nat. In other words, \ptackle can distinguish between a non-matching value for an attribute and a missing attribute. More formally, the semantics of an atomic target $(n, v)$ for a request 
$q = \set{(n_1, v_1), \cdots, (n_k, v_k)}$ is given as:
\[
\sem{(n,v)}(q) =
\begin{cases}
  \match & \text{if $(n,v') \in q$ and $v = v'$}, \\
  \noreqmatch & \text{if $(n,v')  \not\in q$}, \\
  \nomatch & \text{otherwise}.
\end{cases}
\]

More complex targets can be built using the logical operators $\tnot$, for the negation of a target, $\tweakening$, for the optional target (i.e., transform $\noreqmatch$ into $\nomatch$) and $\pand$, for the strong conjunction of targets, interpreted by $\lnot$, $\kweakening$ and $\weakkland$, respectively. Since this set of operators is functionally complete ($\weakkland$ can be built from $\strongklor$ and $\lnot$~\cite{post2012}), any other logical combination can be achieved with them.  

Finally, an {\em authorisation policy} can be defined as single decision, i.e., either $\allow$ (allow) or $\deny$ (deny), a targeted policy $(t, p)$, where $t$ is a target, or a logical composition of two policies, using the operators $\pnot$ for the negation of a policy, $\dbd$ for the deny-by-default of a policy, or $\pand$ for the conjunction of two policies, interpreted by $\lnot$, $\kweakening$ and $\strongkland$, respectively. Here again, these three operators suffice to build any other logical operator. 
\begin{table}[tp]
  \centering
  \caption{\ptackle evaluation}
  \label{tab:example}
  \begin{tabular}{|c||c|c|c|c|}
    \hline
    & $t_1$ & $p_1$ & $t_2$ & $p_2$ \\ \hline \hline
    $\emptyset$ & \,\noreqmatch\, & \,\set{\allow, \deny}\, & \,\noreqmatch\, & \,\set{\allow, \deny}\, \\ \hline
    \set{(\nat, \fr)} & \nomatch & \set{\allow} & \match & \set{\allow} \\ \hline
    \set{(\nat, \at)} & \match & \set{\deny} & \nomatch & \set{\deny} \\ \hline
    \set{(\nat, \fr), (\nat, \at)} & \match & \set{\deny} & \match & \set{\allow}  \\ \hline
  \end{tabular}
\end{table}
Given an access request, the evaluation of a policy returns the set of all possible decisions. The logical operators are therefore extended in a point-wise way, and the evaluation of a targeted policy $(t, p)$ for a request $q$ is given by: 
\[
  \psem{(t, p)}(q) =
  \begin{cases}
    \psem{p}(q) & \text{if $\tsem{t}(q) = \match$}, \\
    \set{\na} & \text{if $\tsem{t}(q) = \nomatch$}, \\
    \set{\na} \cup \psem{p}(q) & \text{otherwise}.
  \end{cases}
\]
where $\na$ represents the not-applicable decision.
For instance, the policy $p_1$ that explicitly denies any access to Austrian citizens and otherwise allows the access can be defined as:  
\begin{atrapenv}
t1 :: (Tatom "nat" "AT")
p1 : Pnot (Pdbd (Pnot (Ptar t1 (Patom Zero))))
\end{atrapenv}
We adopt a declarative syntax, where the double-colon is used for target definition, and a single-colon for policy definition. In the above, the target \atrap{t1} is defined as the atomic target $(\nat, \at)$ with the keyword \atrap{Tatom}, 
\atrap{Patom Zero} represents the atomic policy that always returns $\deny$ (whereas \atrap{One} represents the decision $\allow$), 
\atrap{Ptar t1 (Patom Zero)} is the above policy guarded by the target \atrap{t1}, and thus evaluates to: $\set{\deny}$ if \atrap{t1} evaluates to \match; to 
$\set{\na}$ if \atrap{t1} evaluates to \nomatch; and to $\set{\deny, \na}$ if \atrap{t1} evaluates to \noreqmatch. 
Furthermore, \atrap{Pnot} and \atrap{Pdpd} defines the negation and deny-by-default operators, and therefore the constructor \atrap{Pnot (Pdbd (Pnot x))} acts as an allow-by-default operator, i.e., transforms $\na$ to $\allow$. Similarly, the policy $p_2$ that explicitly authorises any access to French citizens and otherwise denies the access can be defined as:
\begin{atrapenv}
t2 :: (Tatom "nat" "FR")
p2 :  Pdbd (Ptar t2 (Patom One))
\end{atrapenv}
The evaluation of $p_1$ and $p_2$ for four different requests is given in Table~\ref{tab:example}. 
Note that the evaluation might return more than one decision, which can be interpreted as an inconclusive decision. In XACML, such decisions are defined by the {\sf Indeterminate} decision. The way an inconclusive decision is concretely interpreted by the PEP is left to the implementer, and might vary from a risk-advert approach (e.g., any inconclusive decision is interpreted as \deny) to a risk-prone approach (e.g., if \allow is a possible decision, then the PEP allows the request).

\subsection{Maude evaluation}

\ATRAP uses the term rewriting system Maude~\cite{clavel2007all} to model \ptackle and
dynamically generate and evaluate requests for a given policy.  The
syntax for the \ptackle terms in Maude closely resembles the syntax
given above.  Based on this formalisation, we define
\textit{equations} and \textit{rewrite rules} to manipulate the syntax
tree based on pattern matching.

To evaluate a request, we model the operators for \textit{targets},
\textit{policies} and the three valued logic used in \ptackle. The
definition of an operator starts with the keyword \texttt{op},
followed by a pattern that allows parameters at positions marked with
``\texttt{\_}'' (underline), and the signature of the operator after a
``\texttt{:}'' (colon), where the list on the left hand side of
\texttt{->} defines the parameters, and the right hand side the result
type. The definition is completed by a ``\texttt{.}''  (full stop).
To exemplify the notation we give the definitions for decisions and
policy operations:

\begin{maudeenv}
 op ALLOW : -> decision .
 op DENY : -> decision .
 op BOT : -> decision .

 op Patom _ : decision -> policy .
 op Pnot _ : policy -> policy .
 op Pdbd _ : policy -> policy .
 op Pand _ _ : policy policy -> policy .
\end{maudeenv}

The first three lines give parameter-free operators to define the
decisions for $\allow$ (\texttt{ALLOW}), $\deny$ (\texttt{DENY}) and
$\na$ (\texttt{BOT}) respectively.  The decisions are prefixed with
the keyword \texttt{Patom} to form a basic policy, and combined with
\texttt{Pnot}, \texttt{Pdbd} and \texttt{Pand} to form more complex
expressions.  Similar operators exist for the targets, where the basic
element \texttt{Tatom} holds a key-value pair for the
attributes. While these operators capture the structure of the
policies, operators can also be associated with equations to modify
them or evaluate requests.  Equations correspond to operators and are
defined using the keyword \texttt{eq}.  When one of the patterns on
the left of the \texttt{=} matches, it is replaced by the pattern on
the right hand side.

\begin{maudeenv}
 op dbd _ : decision -> decision .
 eq dbd DENY = DENY .
 eq dbd ALLOW = ALLOW .
 eq dbd BOT = DENY .

 op strongand _ _ : decision decision -> decision [assoc comm] .
 eq strongand ALLOW d = d .
 eq strongand DENY d = DENY .
 eq strongand d d = d .

\end{maudeenv}

While the \textit{deny-by-default} operator \texttt{dbd} has only one
parameter and replaces possible \texttt{BOT} values by \texttt{DENY},
the equations can also contain variables which match all possible
patterns for the respective type.  The keywords \texttt{assoc} and
\texttt{comm} declare that the operator for strongand is associative
and commutative respectively, which is considered by Maude in pattern
matching (e.g., we only need \texttt{strongand ALLOW d} and can omit
\texttt{strongand d ALLOW}).  Note that while in general
associative-commutative rewriting is NP complete, Maude supports
effective algorithms for handling the equational rewriting steps for
typical patterns in time proportional to the logarithm of the term
size~\cite{eker2003associative}.  The definitions for equations are
evaluated from top to bottom.  That is, for the equations above,
\texttt{eq strongand d d = d } is only considered if both of the
parameters are \texttt{BOT}.  In addition to the policies, we define
\textit{requests} as sets of \texttt{(key =?  value)} pairs and an
operation \texttt{peval} that evaluates a request on a given policy.

\ATRAP uses this formalisation in three ways: to evaluate a request against a policy, to compute a
counter-example demonstrating that a policy is not resistant, and to search for a proof tree that shows the resistance of a policy.

\section{Policy Resistance}\label{sec:resistance}

To introduce the notion of policy resistance, consider the evaluation of the request \set{(\nat, \fr), (\nat, \at)} with the policy $p_1$, as given in Table~\ref{tab:example}: this request, corresponding to a user with both French and Austrian citizenships, is initially denied; however, if the attribute  (\nat, \at) is ``removed", then the request becomes allowed. 

In general, several reasons can explain the absence of an attribute in a request, such as an error during the transmission of the attributes, the expiration of the attribute certificate, the non-existence of this attribute, etc. In particular, an attribute might be withheld  intentionally (e.g., a user does not want to disclose her address for privacy reasons),  by mistake (e.g., a user is not aware of the fact that it should be disclosed) or maliciously (e.g., a user wants to hide some ``negative'' information). 

In the latter case, the omission of an attribute can be seen as an {\em attribute hiding attack}~\cite{post2012} from a user, trying to gain a better answer by hiding some information. A policy is resistant when it is able to resist to such attacks:

\begin{definition}\label{def:resistance}
A policy $p$ is resistant if, and only if, for any requests $q$ and $q'$, if $q' \subseteq q$ and if $\sem{p}(q') = \set{\allow}$, then $\sem{p}(q) = \set{\allow}$. 
\end{definition}
In other words, if a request $q$ is not allowed, then any sub-request $q' \subseteq q$ is also not allowed. For instance, we can observe that the policy $p_1$ is not resistant, while the policy $p_2$ is.


There are many ways to prove that a policy $p$ is resistant, the most straight-forward one being to exhaustively check any pair of requests $q, q'$ such that $q' \subseteq q$. We describe an implementation of this approach in Section~\ref{sec:counterexample} 
using Maude, together with the notion of a normal form for requests, allowing us to reduce the set of requests to check. 

In some cases, we can also use the {\em structure} of the policy to prove its resistance. For instance, the policy 
\atrap{Ptar t (Patom Zero)} clearly evaluates, for any request, either to $\set{\deny}$ or to $\set{\deny, \bot}$, regardless of the definition of \atrap{t}, and is therefore trivially resistant. Generalising this example, we can observe (and formally prove) that if a policy cannot return $\allow$, then it is resistant. Furthermore: 
\begin{itemize}
\item if \atrap{p} cannot return $\allow$, then \atrap{Ptar t p} cannot return $\allow$, for any \atrap{t};
\item if \atrap{p} cannot return $\allow$, then \atrap{Pdbd p} cannot return $\allow$; 
\item if \atrap{p} cannot return $\allow$, then \atrap{Pand p p1} and \atrap{Pand p1 p} cannot return $\allow$; 
\item if \atrap{p} cannot return $\allow$, then \atrap{Pnot p} cannot return $\deny$;
\item if \atrap{p} cannot return $\deny$, then \atrap{Pnot p} cannot return $\allow$;
\item if \atrap{p} cannot return $\deny$, then \atrap{Ptar t p} cannot return \deny. 
\end{itemize}
In other words, it might be possible to prove that a policy is resistant simply by inspecting its structure, without checking each possible pair of requests. As another example, a policy without any target clearly evaluates identically for any request (since all requests are equally applicable), and therefore is resistant. 

\medskip

In addition, the notion of weak-monotonicity is introduced in~\cite{post2012}, such that, intuitively speaking, a target is weakly monotonic if removing information from a request lowers the evaluation of the target. 

\begin{definition}
A target $t$ is \emph{weakly monotonic} if for all requests $q$ and for every $q' \subseteq q$,  $\sem{t}(q') \preccurlyeq \sem{t}(q)$, where $\preccurlyeq$ is the reflexive closure of $\noreqmatch \prec \nomatch \prec \match$.  A policy $p$ is weakly monotonic if and only if every target in $p$ is weakly monotonic. 
\end{definition}

Any atomic target $(n, v)$ is weakly-monotonic, and the operators $\topt$ and $\tand$ preserve the weak-monotonicity~\cite{post2012}. 
In other words, any policy whose targets are built without the operator $\tnot$ is weakly monotonic. Moreover, as a direct result from Theorem 6 of~\cite{post2012} (which is recalled in Appendix~\ref{thm:weak-monotonicity}), we have: 
\begin{itemize}
\item If \atrap{p} is weakly monotonic and built without $\dbd$, then it is resistant; 
\item If \atrap{p} is weakly monotonic and built without $\pnot$, then it is resistant; 
\end{itemize}
Finally, the notion of resistance can be proved in a compositional way: 
\begin{itemize}
\item If \atrap{p} is resistant, then \atrap{Pdbd p} is resistant; 
\item If \atrap{p1} and \atrap{p2} are resistant, then \atrap{Pand p1 p2} is resistant. 
\end{itemize}
These rules make it possible to prove the resistance of a conjunctive policy in a different way for each sub-policy. Clearly, all the rules presented in this section  are only implications, and a policy might be resistant even though it does not satisfy any of them. 
We show in Section~\ref{sec:certification} how \ATRAP can use these rules, together with their encoding in Isabelle and Maude, to automatically build a proof of resistance. 

\begin{remark}
  A  (strongly) monotonic policy, as defined in~\cite{post2012} is also trivially resistant. However, in order to prove the (strong) monotonicity of an atomic target, attributes must be assumed to be {\em compact}, i.e., either all the values of an attribute are given, or none are. For instance, the compactness of the attribute $\nat$ would mean that a user can either hide all of her nationalities, or none of them, but cannot only hide one. Hence, proving resistance using (strong) monotonicity requires the assumption of compactness from the environment, whereas we aim here at generating {\em complete} proofs, i.e., without assumption. The integration of such assumptions when all other strategies have failed is left for future work. 
\end{remark}

\section{Search for non-resistance}
\label{sec:counterexample}

We use the \ptackle encoding in Maude to automatically search for
possible counter-examples to resistance.  A naive approach for a policy
$p$ would simply consist in checking any two requests $q$ and $q'$,
such that $q' \subseteq q$ and $\sem{p}(q') = \set{\allow}$, in order
to ensure that $\sem{p}(q) = \set{\allow}$. However, the set of all
possible requests can potentially be very large. For instance, in our
previous example, focusing only on the \nat attribute, the United
Nations Organisation currently counts 193
members\footnote{\url{http://www.un.org/depts/dhl/unms/whatisms.shtml#states}},
meaning there are 193 pairs $(\nat, v)$ possible, and thus
$\mathbf{card}(\Q) = 2^{193}$.

In order to simplify the search, we first introduce the {\em normal form} of a request for a given policy. We then describe our correct and complete search for counter-examples in Maude, and finally we present some experimental results. 

\subsection{Normal form of requests}
Intuitively speaking, the evaluation of a request against a policy mostly depends on whether the request contains the atomic targets present in the policy: given an attribute $n$, all pairs $(n, v)$ that do not explicitly appear in the policy are evaluated in the same way. For instance, consider the policy $p_1$: it is clear that the pairs $(\nat, \fr)$, $(\nat, \de)$ or $(\nat, \uk)$ are evaluated similarly. 

Hence, given a policy $p$, we write $\Att(p)$ for the set of atomic targets appearing in $p$. For instance, 
$\Att(p_1) = \set{(\nat, \at)}$ and $\Att(p_2) = \set{(\nat, \fr)}$. Given an attribute $n$ and a policy, we write 
$\fv(p, n)$ for a fresh value of $n$ with respect to $p$, i.e., a value such that $(n, \fv(p, n)) \not\in \Att(p)$. If $p$ explicitly mentions all possible values for $n$, we define $\fv(p, n)$ to return a random value for $n$. Finally, the normal form of a request $q$ for a policy $p$ is given by keeping all pairs $(n, v)$ that appear both in $q$ and $\Att(p)$, and replacing any $(n, v)$ in $q$ that does not appear in $\Att(p)$ by $(n, \fv(p, n))$. More formally: 
\begin{definition}
  The normal form of a request $q$ is given by:
  \[
  \nf_p(q) = (q \cap \Att(p)) \cup \set{(n, \fv(p, n)) \mid \exists v\,\, (n, v) \not \in \Att(p) \land (n, v) \in q }
  \]
\end{definition}
Given a set of requests $\Q$ and a policy $p$, we write $\NF_p(\Q) = \set{\nf_p(q) \mid q \in \Q}$. In the following, we omit the subscript when $p$ is clear from context. For instance, the set of requests in normal form for the policy $p_1$ is given by: 
\[
\NF_{p_1}(\Q) = \set{\emptyset, \set{(\nat, \at)}, \set{(\nat, \nv)},\set{(\nat, \nv), (\nat, \at)}}
\]
where $\nv$ represents the fresh value for the attribute $\nat$, i.e., $\nv=\fv(p_1, \nat)$.

This notion of normal is consistent both with policy evaluation and request inclusion (the proofs can be found in Appendix~\ref{sec:proof}). 
\begin{prop}\label{thm:nf}
  For any policy $p$ and any request $q$, $\sem{p}(q) = \sem{p}(\nf(q))$.
\end{prop}
\begin{prop}\label{thm:inclusion}
Given any requests $q$ and $q'$, if $q' \subseteq q$, then $\nf(q') \subseteq \nf(q)$. 
\end{prop}

\begin{remark}
The simplicity of the normal form comes from the fact that \ptackle does not allow for 
complex atomic targets, such as the comparison of attribute values. For instance, one cannot directly write the atomic target stating that \textbf{age} must be lower than 18, and must instead generate the disjunction of all atomic target for \textbf{age} between 0 and 18. 
Note however that, as long as attributes have a finite domain, all possible targets can be defined, hence \ptackle can be seen here as a low-level language, designed for policy analysis. The design of richer atomic targets is planned for future work. 
\end{remark}

\subsection{Search for counter-examples}
\newcommand{\card}[1]{\ensuremath{ \left\vert{#1}\right\vert}}
As a consequence of Proposition~\ref{thm:nf} and Proposition~\ref{thm:inclusion}, the resistance of a policy can be decided only by looking at the set of requests in normal form. 

\begin{prop}\label{thm:restriction}
A policy $p$ is resistant if, and only if, given $q$ and $q'$ in $\NF_p(\Q)$, 
if $q' \subseteq q$ and if $\sem{p}(q') = \set{\allow}$, then $\sem{p}(q) = \set{\allow}$. 
\end{prop}
In other words, we can restrict our attention to $\NF_p(\Q)$, whose size is bounded by $2^{\card{\Att(p)} + n}$, where $n$ stands for the number of attributes, instead of $\Q$, whose size is $2^N$, where $N$ is the number of all possible attribute name-values pairs.  
Finally, it is worth observing that, in order for a policy $p$ to be resistant, it is enough 
to check, for any allowed request $q$, whether removing any pair attribute value changes
the decision. More formally: 
\begin{prop}\label{thm:step}
  A policy $p$ is resistant if, and only if, for any request $q$ such that $\sem{p}(q) \neq \set{\allow}$, if $\sem{p}(q \setminus \set{(n, v)}) \neq \set{\allow}$, for any attribute $n$ and any value $v$. 
\end{prop}

Combining Propositions~\ref{thm:step} and~\ref{thm:restriction}, we conclude that to find a
counter-example to resistance, we only need to check all pairs $(q,
q')$, where $q, q' \in \NF_p(\Q)$ and $q = q' \cup \set{(n, v)}$.
Proposition~\ref{thm:step} allows us to reduce the number of
comparisons needed for the search for counter-examples from an upper
bound of $2^{\card{\Att(p)} + n}\times 2^{\card{\Att(p)}+n}$ (comparison of all subsets) to $2^{\card{\Att(p)}+n} \times
(\card{\Att(p)} + n)$, i.e., where each subset needs to be checked against at most $\card{\Att(p)} + n$ direct subsets.



The seach for counter-examples is performed by generating and
evaluating the largest possible request $q_m = \Att(p) \cup \set{(n,
  \fv(n)) \mid (n, v) \in \Att(p)}$, and systematically removing
attributes to see if a reduction of a request (hiding of an attribute)
can lead to an increase of access in the policy.  This manipulation of
requests is performed using \textit{rewrite rules}, which are defined
similarly to equations, but in contrast to them are not evaluated
deterministically, i.e., may be executed whenever the left hand side
pattern matches.

The Maude command for the search has the following form:
\begin{maudeenv}
search sres(bldevallist (policy Requests Defs DecsList ) )  =>* error ( x:DecsList ) . 
\end{maudeenv}
where \maude{policy} is the ID of a top level policy to check, the
set \maude{Requests} is the maximal request in normal form, and
\maude{Defs} are the policy definitions.  The operator
\maude{bldevallist} repeatedly removes attribues from the request,
evaluates it with respect to the policy, and stores the result in
\maude{DecsList}.  The operator \maude{sres} traverses the list and
searches for pairs that violate resistance, in which case the
violating request pair is wrapped into an \maude{error} operator.
Removing an element from the set \maude{Request} is nondeterministic,
and thus may generate different lists of decisions.  The maude command
\maude{search} systematically explores all the possible outcomes and
returns those that match the search command, resp. can be rewritten to
an \maude{error} label. For instance, when analysing the policy $p_1$, \ATRAP outputs:
\begin{atrapenv}
Counter-example #1
["nat" =? "new_value"]:                    [ALLOW]
["nat" =? "AT", "nat" =? "new_value"]:  [DENY]
\end{atrapenv}

\subsection{Experimental Results}
\label{sec:exp_ce}
  \begin{figure}[tp]
\vspace{-15pt}
    \caption{Automatic search for counter-examples for $P\langle 4, 4, 4, 4, 300\rangle$}
    \vspace{10pt}
      

    \includegraphics{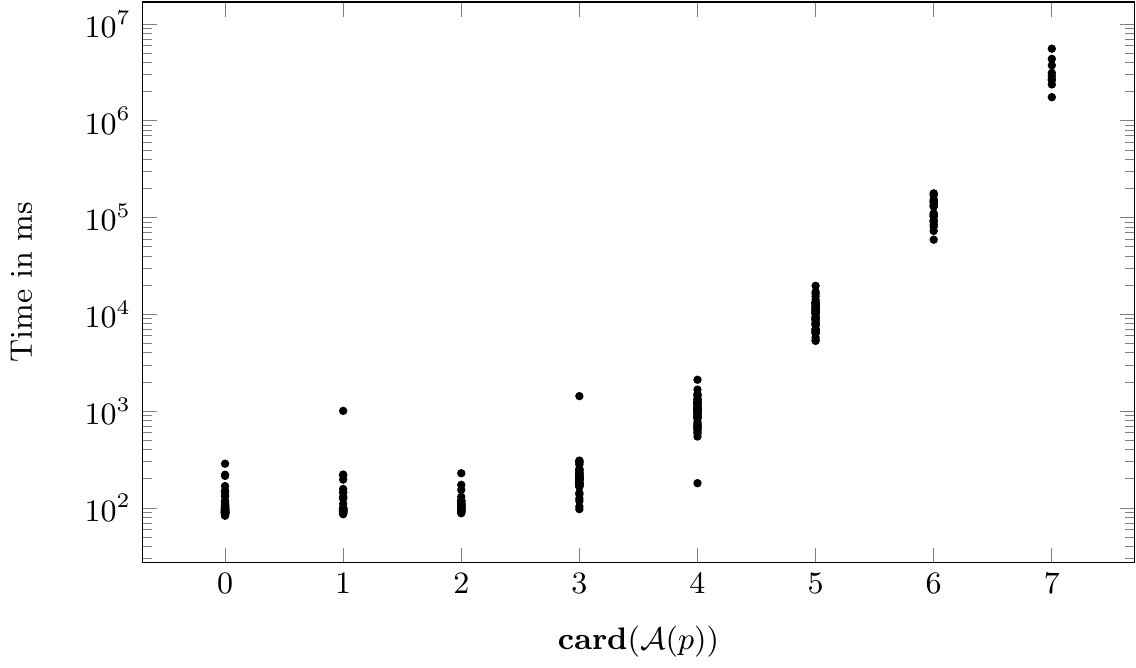}
    \label{fig:countersearch}
\vspace{-15pt}
  \end{figure}

We experiment the search for counter-examples by randomly generating some policies, and executing the search for each policy.  
We write $P\langle m, n, k, l, r\rangle$ for a set of $r$ random policies, such that $m$ stands for the maximal height of each policy, $n$ for the maximal width of each target, $k$ for the number of attributes and $l$ for the number of values for each attribute.




Figure~\ref{fig:countersearch} represents the execution time (on 2 GHz Intel core i7 with 8GB of RAM) for the search of counter-examples 
for each $p \in P\langle 4, 4, 4, 4, 300\rangle$, indexed by $\Att(p)$ and with a logarithmic scale for the execution time. Over the 300 policies, 252 are resistant, and this ratio of 0.85 is consistent with other experiments, and seems to be independent of the policy dimensions. Note that the times shown in this Figure are for complete searches, i.e., searches finding {\em all} possible counter-examples. Hence, the time required to analyse a policy is the same whether the policy is resistant or not. As expected from the theoretical analysis of the previous section, the search for counter is exponential in the size of $\Att(p)$.

\section{Certification of resistance}
\label{sec:certification}

If no counter-example is found, \ATRAP tries to build a proof of
resistance of the policy, following a proof-obligation discharge
approach, where Maude is used to automatically generate a proof that
can be checked in Isabelle~\cite{DBLP:books/sp/NipkowPW02,Wenzel07theisabelle/isar}.  To reach this goal, \ATRAP encodes the
policy and the deduction mechanism in rewrite logic and calls Maude to
perform a search for the proof.  The result is parsed into Java
classes and mapped into a corresponding proof in Isabelle, which is
then called to validate the proof. It is worth mentioning that the
encoding in Maude of the deduction mechanism is independent to the
counter-examples search described in the previous section, although
both techniques share the encoding of \ptackle.


Our approach revolves around two main entities: proof obligation and
proof technique.  A {\em proof obligation} corresponds to a goal,
i.e., to a property that we want to prove.  The top-level goal is a
proof of resistance, which in turn may require further sub-goals like
weak-monotonicity to form a complete proof tree. A {\em proof
  technique} describes a method to discharge a proof obligation, using
some rules described in Section~\ref{sec:resistance}.

Rather than encoding the proofs manually in Isabelle, we use Maude to
search for viable proof trees that are then encoded and checked in
Isabelle.  To facilitate this goal, each proof obligation and
technique have corresponding facets in Maude, Isabelle and Java.  In
Maude, the facet of a proof obligation is represented as an operator
(e.g., \maude{isResistant}$\!$) that takes a policy or target ID and a
proof technique. In Java this is implemented in form of public
classes, which have a field \java{statement} corresponding to the
Isabelle definition of the proof obligation. For instance, the field
\java{statement} of the class \java{ResistanceProof} is equal to
\java{resistant p q1 q}, where \java{p} is the name of the policy over
which this class is defined and where \java{q} and \java{q1} are the
universally quantified variables representing the variables $q$ and
$q'$ of Definition~\ref{def:resistance}.  The Isabelle facet of a
proof obligation is then a definition over either a target
or a policy. For instance, the definition of resistance in Isabelle is
given as:
\begin{isaenv}
definition resistant :: "policy \<Rightarrow> request \<Rightarrow> request \<Rightarrow> bool" where 
"resistant p q1 q \<equiv> (set q1)\<subseteq>(set q) \<longrightarrow> peval p q1 = {One} \<longrightarrow>  peval p q = {One}"
\end{isaenv}

The Maude facet of a proof technique is an operator defined over
policies and/or targets.  For instance, let $r_1$ be the rule stating
that a policy is resistant if it is weakly-monotonic and without the
$\pnot$ operator. Note that for technical reasons, we also need to
impose that a policy is well-formed, i.e., each atomic policy is
either \allow or \deny, and not \na.  This rule is represented in
Maude as the operator \maude{ResWFWMWNProof}. A proof technique is
implemented in Java as a class local to that corresponding to the
proof obligation. For instance, in order to define the rule $r_1$, the
class \java{ResistanceProof} comes with a local class
\java{WeakMonotonicityWithoutNotWF}.  This class is defined with three
fields: \java{WellFormedProof}, \java{WithoutNotProof},
\java{WeakMonotonicPolicyProof}, each being a public class. In other
words, in order to use the rule $r_1$, one must first exhibit a proof
that the policy is well-formed, a proof that it is built without the
$\pnot$ constructor, and a proof that it is weakly-monotonic.
Finally, this structure is mapped to Isabelle, where the proof
technique for a proof obligation is a proven lemma whose goal is that
proof obligation.  For instance, the lemma corresponding to rule $r_1$
is defined in Isabelle as:
\begin{isaenv}
lemma weak_monotonic_without_not_resistant :  
"well_formed_policy p \<Longrightarrow> weak_pmonotonic p q1 q \<Longrightarrow> policy_without_not p 
  \<Longrightarrow> resistant p q1 q"
\end{isaenv}

To generate the proof tree, each rule of Section~\ref{sec:resistance} is modelled in Maude in a basic,
compositional form.  Starting from the initial goal of proving
resistance of a policy $p$, the proof generation then follows the
structure of the policy to generate new proof obligations for
sub-proofs according to the components and properties of $p$.

While binary operators like \atrap{Pand} trigger sub-proofs for both
operands, the proof generation is also guided by properties and
preconditions of the lemma to apply.  To show, e.g., resistance using
the rule $r_1$, we need to
establish well-formedness, weak-monotonicity, and check that the
policy does not use $\pnot$.  Well-formedness and use of $\pnot$ can
easily be checked by Maude doing a syntactic check.  Only if both
conditions are fulfilled, a proof for weak monotonicity is
instantiated. When all conditions for a proof are fulfilled, the \textit{proof
  obligation} is replaced by a description of the actual proof. For instance, 
the rewrite rule corresponding to the rule $r_1$ is given as:
\begin{maudeenv}
   genproof(isResistant(p, noproof), 
            isWF(p, pr1), isWN(p, pr2), isWM(p, pr3),pis | defs )
=> genproof(isResistant(p, ResWFWMWNProof(p)), 
            isWF(p, pr1), isWN(p, pr2), isWM(p, pr3), pis | defs)
\end{maudeenv}
where \maude{noproof} indicates that the proof obligation was not
fulfilled yet, and \maude{pr1}, \maude{pr2}, and \maude{pr3} are
previously generated subproofs.  The variables \maude{pis} and
\maude{defs} hold the available sub-proofs and policy definitions
respectively.  This approach also allows for manual intervention by
the user by supplying the generation mechanism by external information
about the system or predefining proof obligations with their
respective techniques.  



  

\begin{figure}[tp]
  \centering
  \caption{Generated Isabelle Proof}
  \label{fig:proof}
\begin{isaenv}
theory p2 imports atrap begin
definition t2 :: target where "t2 = (Tatom ''nat'' ''FR'')"
definition pone :: policy where "pone = Patom One"
definition pt :: policy where "pt = Ptar t2 pone"
definition p2 :: policy where "p2 = Pdbd pt"

lemma "resistant p2 q1 q" proof -
  have wf: "well_formed_policy p2" 
    by (simp add: p2_def pt_def pone_def t2_def)
  have without_not: "policy_without_not p2" 
    by (simp add: p2_def pt_def pone_def t2_def)
  have weak_monotonic: "weak_pmonotonic p2 q1 q" proof -
    have wm_p: "weak_pmonotonic pt q1 q" proof -
        have wm_p: "weak_pmonotonic pone q1 q" by (simp add: pone_def)
        have wm_t: "weak_tmonotonic t2 q1 q" 
          by (simp add: tatom_weak_monotonic t2_def)
        from wm_p wm_t show ?thesis by (simp add:pt_def) qed     
    from wm_p show ?thesis by (simp add:p2_def) qed
  from wf without_not weak_monotonic show ?thesis
    by (insert weak_monotonic_without_not_resistant [of p2 q1 q], simp)
qed
\end{isaenv}
\end{figure}

\subsubsection*{Running Example.}
The policy $p_2$ is automatically proven to be resistant by \ATRAP, which executes the following Maude command (where the policy 
\atrap{pt} is introduced as an intermediary step):  
\begin{maudeenv}
rew genproof(isResistant(P "p2", noproof) | (T "t2"::(Tatom "nat" "FR"), 
  P "pone"=Patom ALLOW, P "pt"=Ptar T "t2" P "pone", P "p2"=Pdbd P "pt")). 
\end{maudeenv}
This proof-obligation is automatically discharged using the rules described above, and the following proof-obligation is returned:
\begin{maudeenv}
isResistant(P "p2", ResWFWMWNProof(P "p2")), isWF(P "p2", WFBFProof(P "p2")), 
isWN(P "p2", WNBFProof(P "p2")), isWM(P "p2", WMPdbd(P "p2",P "pt")), 
isWM(P "pt", WMwithPtar(P "pt",T "t2",P  "pone")), 
isWM(T "t2" ,WMwithTAtom(T "t2")), isWM(P "pone",  WMwithPAtom(P "pone"))) 
\end{maudeenv}
Informally, this proof can be read as follows: \atrap{p2} is resistant, since it is well-formed, weakly-monotonic and without-not; 
\atrap{p2} is well-formed which can checked by ``brute-force'', i.e., by checking the definition of \atrap{p2}; 
\atrap{p2} is without the $\pnot$ operator, which can also be checked by ``brute-force'';  
\atrap{p2} is weakly-monotonic, since it is the deny-by-default of the weakly-monotonic policy \atrap{pt};
\atrap{pt} is weakly-monotonic, since it is the composition of the weakly-monotonic target \atrap{t2} and 
of the weakly-monotonic policy \atrap{pone}; \atrap{t2} is weakly-monotonic, since it is atomic; and \atrap{pone} is weakly-monotonic since it is atomic. 

\ATRAP parses this Maude proof-obligation, and using the Java mechanism described above, the following Isabelle theory is  automatically generated. This theory is built on the logic \isa{atrap.thy}, which includes the definition of the three-valued logic, the definition of \ptackle, and the lemmas corresponding to the different rules described in Section~\ref{sec:resistance}. 

Fig.~\ref{fig:proof} presents the generated proof for the running example. For the sake of compactness, we do not go through Isabelle/Isar's syntax, but intuitively, the structure of the proof follows the informal description given above, and the tactics used are limited to the simplification tactic (which unfolds the definition of the entities involved in the proof), and the insertion of existing lemmas. It is worth observing that the generated proof is human readable, and is structurally very close to the corresponding mathematical proof. We believe this aspect to be particularly important as a security designer is not necessarily an expert in proof techniques, and \ATRAP provides a high-level proof, without having to rely blindly on a verification tool.

\subsection{Experimental results}
\label{sec:exp_pg}
We now evaluate the performance of \ATRAP for the generation of resistance proof.
The complexity of the proof generation mostly depends on the number of constructors of a policy $p$, which we refer to by $size(p)$, since 
the number of applicable rules directly depends on that number. 
\begin{figure}[tp]
  \caption{Automatic search for proof for $P\langle 8, 3, 3, 3, 500\rangle$}
  \vspace{10pt}

    \includegraphics{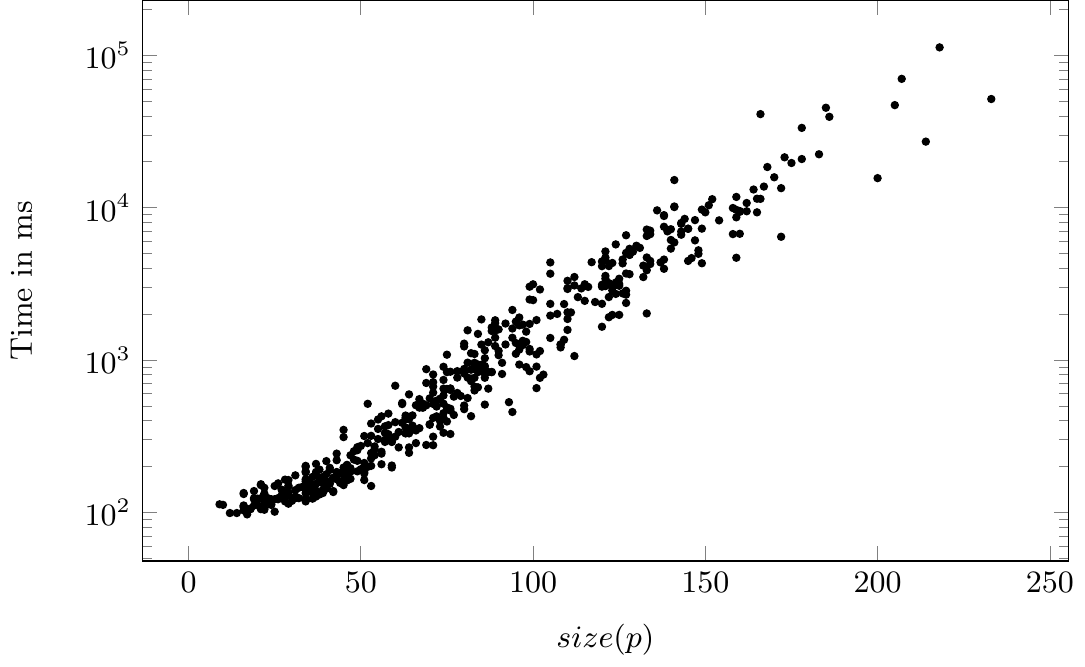}
  \label{fig:proofsearch}
\end{figure}
Figure~\ref{fig:proofsearch} shows some evaluation times for $P\langle 8, 3, 3, 3, 500\rangle$, where the $y$ axis is logarithmic. As 
expected, the complexity of the proof search is exponential in $size(p)$, making the proof search most of the time faster than the search for counter-examples (the few exceptions to that rule come from policies $p$ with large sizes, but small $\Att(p)$, i.e., policies where a same target is repeated a large number of times). 

However, although usually faster than the counter-examples search, the accuracy of the proof generation decreases with $size(p)$. 
For instance, writing $T_n$ for $P\langle n, n, 2, 2, 1000\rangle$, for $T_1$, we generated the proof for all of the 957 resistant policies, this ratio falls to $905 / 920 \approx 0.98$ for $T_2$, to $762 / 883 \approx 0.86$ for $T_3$, to $ 659 / 864 \approx 0.76$ for $T_4$, to $558 / 852 \approx 0.66$ for $T_5$, to $503 / 861 \approx 0.58$ for $T_6$, etc.

\medskip


Theses results, together with those of Section~\ref{sec:exp_ce}, should be seen as a validation of the \ATRAP approach, rather than a ``real-world'' characterisation of policy resistance. Indeed, the policies analysed are randomly generated, and therefore the samples do not necessarily represent policies defined in a concrete context. In particular, it is not necessarily the case that about 85\% of the policies enforced in existing information systems are resistant. Similarly, even though the accuracy of the proof search decreases with the number of constructors used, a very large policy whose targets consist only of conjunctions of atomic targets, and using only the operator $\pand$, could be easily proved resistant.

\section{Conclusion - Future work}
\label{sec:conclusion}

This paper presents the tool \ATRAP, which, given a \ptackle policy, is capable of positively answer the three questions stated in the Introduction, i.e., can automatically detect the resistance of a policy by exhibiting a counter-example when it is not resistant, and, in some cases, by generating an Isabelle proof when it is. The different mechanisms are illustrated with two simple policies. 

Using Proposition~\ref{thm:restriction}, we are able to limit the number of requests to evaluate when searching a counter-example, while maintaining the correctness and completeness of the approach. We have also implemented a collection of proof techniques allowing to prove efficiently the resistance of a policy, although large policies might fail to be proven automatically. However, the rules presented in Section~\ref{sec:resistance} can also be seen as a policy construction guide, since a policy built using only those rules is resistant, by construction. 

An interesting lead to explore for future work is to increase the interactivity between the search for counter-examples and the proof generation, by leveraging the different complexity of each approach. More precisely, we should take advantage of Maude to rewrite a policy to an equivalent one, such that the latter policy could be proven to be resistant more easily. This leads to the question of the existence of a normal form for policies, such that one could build a complete collection of resistance rules, i.e., if a policy in normal form is resistant, then there exists a set of structural proofs to prove it. At this stage, this remains as an open question. 

We however believe that one of the strengths of our approach is its flexibility, and rules can be incrementally added. In order to so, one needs to provide the corresponding rule in Maude, together with Isabelle lemma and the Java class linking the two. Clearly, an interesting future work would consist in formalising this extendability, by having an explicit, abstract notion of rule, with multiple facets, i.e., one facet for Maude, one for Isabelle, one for Java. We could also extend our approach to other properties of access control policies, for instance by relying on the underlying properties of the rewrite system~\cite{Bertolissi:2013:AAR:2428116.2428125}. 

Another relevant problem is the one of fixing a policy, in order to transform a non-resistant policy into a resistant one. This problem raises the question of policy ``closeness'', i.e., given a non-resistant policy, it is not enough to create a resistant one, we also need to ensure that the new policy is close enough to the original one. It is worth noting that \ATRAP can generate partial proof of resistance, i.e., even if the whole policy is not resistant, it might identify some sub-policies that are, which might be helpful to fix a given policy. 


Finally, as we mentioned, the current version of \ptackle can be seen as a low level language, and it would therefore be worth interfacing a ``higher'' level language with \ptackle, in order to analyse more complex policies, for instance include relational targets. 
XACML 3.0 would be a good candidate for such an extension, since both languages are attribute and target-based. In general, we plan to release \ATRAP as an open-source software, and to generally optimise the searches for counter-examples and proofs. In this regard, it would be worth looking at parallel/distributed computation, since the evaluation of each pair of requests $(q, q \setminus \set{(n, v)}$ can be performed independently. 

\medskip
\noindent
{\bf Acknowledgements.} The authors would like to thank Jason Crampton for valuable discussions about \ptackle and policy resistance. 

\bibliographystyle{abbrv}
\bibliography{fbibli}

\begin{thebibliography}{10}

\bibitem{Alberti:2011:ESA:1966913.1966935}
F.~Alberti, A.~Armando, and S.~Ranise.
\newblock Efficient symbolic automated analysis of administrative
  attribute-based rbac-policies.
\newblock In {\em Proceedings of the 6th ACM Symposium on Information, Computer
  and Communications Security}, ASIACCS '11, pages 165--175, New York, NY, USA,
  2011. ACM.

\bibitem{DBLP:journals/jcs/ArmandoR12}
A.~Armando and S.~Ranise.
\newblock Scalable automated symbolic analysis of administrative role-based
  access control policies by smt solving.
\newblock {\em Journal of Computer Security}, 20(4):309--352, 2012.

\bibitem{Barker:2006:TRA:2090485.2090498}
S.~Barker and M.~Fern\'{a}ndez.
\newblock Term rewriting for access control.
\newblock In {\em Proceedings of the 20th IFIP WG 11.3 working conference on
  Data and Applications Security}, DBSEC'06, pages 179--193, Berlin,
  Heidelberg, 2006. Springer-Verlag.

\bibitem{Bertolissi:2013:AAR:2428116.2428125}
C.~Bertolissi and W.~Uttha.
\newblock Automated analysis of rule-based access control policies.
\newblock In {\em Proceedings of the 7th workshop on Programming languages
  meets program verification}, PLPV '13, pages 47--56, New York, NY, USA, 2013.
  ACM.

\bibitem{bona:alge02}
P.~Bonatti, S.~{De Capitani Di Vimercati}, and P.~Samarati.
\newblock An algebra for composing access control policies.
\newblock 5(1):1--35, 2002.

\bibitem{Brucker:2011:AMT:1998441.1998461}
A.~D. Brucker, L.~Br\"{u}gger, P.~Kearney, and B.~Wolff.
\newblock An approach to modular and testable security models of real-world
  health-care applications.
\newblock In {\em Proceedings of the 16th ACM symposium on Access control
  models and technologies}, SACMAT '11, pages 133--142, New York, NY, USA,
  2011. ACM.

\bibitem{DBLP:journals/tissec/BrunsH11}
G.~Bruns and M.~Huth.
\newblock Access control via {Belnap} logic: Intuitive, expressive, and
  analyzable policy composition.
\newblock {\em ACM Transactions on Information and System Security}, 14(1):9,
  2011.

\bibitem{clavel2007all}
M.~Clavel, F.~Dur{\'a}n, S.~Eker, P.~Lincoln, N.~Mart{\'\i}-Oliet, J.~Meseguer,
  and C.~Talcott.
\newblock {\em All about {Maude} - a high-performance logical framework: how to
  specify, program and verify systems in rewriting logic}, volume 4350 of {\em
  LNCS}.
\newblock Springer, 2007.

\bibitem{CH10}
J.~Crampton and M.~Huth.
\newblock An authorization framework resilient to policy evaluation failures.
\newblock In D.~Gritzalis, B.~Preneel, and M.~Theoharidou, editors, {\em
  ESORICS}, volume 6345 of {\em Lecture Notes in Computer Science}, pages
  472--487. Springer, 2010.

\bibitem{post2012}
J.~Crampton and C.~Morisset.
\newblock {PTaCL}: A language for attribute-based access control in open
  systems.
\newblock In P.~Degano and J.~D. Guttman, editors, {\em POST}, volume 7215 of
  {\em Lecture Notes in Computer Science}, pages 390--409. Springer, 2012.

\bibitem{Dougherty:2007:MAC:2393847.2393899}
D.~J. Dougherty, C.~Kirchner, H.~Kirchner, and A.~S. De~Oliveira.
\newblock Modular access control via strategic rewriting.
\newblock In {\em Proceedings of the 12th European conference on Research in
  Computer Security}, ESORICS'07, pages 578--593, Berlin, Heidelberg, 2007.
  Springer-Verlag.

\bibitem{eker2003associative}
S.~Eker.
\newblock Associative-commutative rewriting on large terms.
\newblock In {\em Rewriting Techniques and Applications}, pages 14--29.
  Springer, 2003.

\bibitem{Griesmayer:2013:FAC:2443461.2443470}
A.~Griesmayer, Z.~Liu, C.~Morisset, and S.~Wang.
\newblock A framework for automated and certified refinement steps.
\newblock {\em Innov. Syst. Softw. Eng.}, 9(1):3--16, Mar. 2013.

\bibitem{DBLP:conf/isw/GuelevRS04}
D.~P. Guelev, M.~Ryan, and P.-Y. Schobbens.
\newblock Model-checking access control policies.
\newblock In K.~Zhang and Y.~Zheng, editors, {\em ISC}, volume 3225 of {\em
  Lecture Notes in Computer Science}, pages 219--230. Springer, 2004.

\bibitem{Hughes:2008:AVA:1459278.1459282}
G.~Hughes and T.~Bultan.
\newblock Automated verification of access control policies using a sat solver.
\newblock {\em Int. J. Softw. Tools Technol. Transf.}, 10(6):503--520, Oct.
  2008.

\bibitem{Jobe62}
W.~Jobe.
\newblock Functional completeness and canonical forms in many-valued logics.
\newblock {\em Journal of Symbolic Logic}, 27(4):409--422, 1962.

\bibitem{Kirchner:2009:ARA:1519544.1519756}
C.~Kirchner, H.~Kirchner, and A.~S. de~Oliveira.
\newblock Analysis of rewrite-based access control policies.
\newblock {\em Electron. Notes Theor. Comput. Sci.}, 234:55--75, Mar. 2009.

\bibitem{kleene:intr50}
S.~Kleene.
\newblock {\em Introduction to Metamathematics}.
\newblock D. Van Nostrand, Princeton, NJ, 1950.

\bibitem{DBLP:books/sp/NipkowPW02}
T.~Nipkow, L.~C. Paulson, and M.~Wenzel.
\newblock {\em Isabelle/HOL - A Proof Assistant for Higher-Order Logic}, volume
  2283 of {\em Lecture Notes in Computer Science}.
\newblock Springer, 2002.

\bibitem{xacml2.0}
OASIS.
\newblock {\em eXtensible Access Control Markup Language (XACML) Version 2.0},
  2005.
\newblock Committee Specification.

\bibitem{XACML3}
OASIS.
\newblock {\em eXtensible Access Control Markup Language (XACML) Version 3.0},
  2010.
\newblock Committee Specification 01.

\bibitem{Rao:2009:AFI:1542207.1542218}
P.~Rao, D.~Lin, E.~Bertino, N.~Li, and J.~Lobo.
\newblock An algebra for fine-grained integration of {XACML} policies.
\newblock In {\em Proceedings of the 14th ACM Symposium on Access Control
  Models and Technologies}, pages 63--72, New York, NY, USA, 2009. ACM.

\bibitem{confsacmatTschantzK06}
M.~Tschantz and S.~Krishnamurthi.
\newblock Towards reasonability properties for access-control policy languages.
\newblock In D.~Ferraiolo and I.~Ray, editors, {\em {SACMAT} 2006,11th {ACM}
  Symposium on Access Control Models and Technologies, Proceedings}, pages
  160--169. ACM, 2006.

\bibitem{Wenzel07theisabelle/isar}
M.~Wenzel.
\newblock The isabelle/isar reference manual, 2007.

\bibitem{wije:prop03}
D.~Wijesekera and S.~Jajodia.
\newblock A propositional policy algebra for access control.
\newblock 6(2):286--235, 2003.

\bibitem{Zhang:2008:SVA:1370684.1370685}
N.~Zhang, M.~Ryan, and D.~P. Guelev.
\newblock Synthesising verified access control systems through model checking.
\newblock {\em J. Comput. Secur.}, 16(1):1--61, Jan. 2008.

\end{thebibliography}

\appendix
\section{Original \ptackle theorem}
\label{app:post}

For the ease of reading of this paper, we restate here the theorem about the weak-monotonicity of~\cite{post2012}. 
\begin{Thm}\label{thm:weak-monotonicity}
Let $p$ be a policy whose policy tree contains \emph{weakly monotonic} targets $t_1,\dots,t_k$ and let $q$ be a request.
  \begin{compactenum}
    \item If $p$ is constructed from the operators $\pnot$ and $\pand$, then for any $q' \subseteq q$, if $\sem{p}(q') = \set{d}$, with $d \in \set{\allow, \deny}$, then $\sem{p}(q) = \sem{p}(q')$.
    \item If $p$ is constructed from the operators $\dbd$ and $\pand$, then for any $q' \subseteq q$, if $\sem{p}(q') = \set{\allow}$, then \mbox{$\sem{p}(q) =  \set{\allow}$}.
  \end{compactenum}
\end{Thm}

\section{Proofs}
\label{sec:proof}

\noindent \textbf{Proposition~\ref{thm:nf}}\textit{
  For any policy $p$ and any request $q$, $\sem{p}(q) = \sem{p}(\nf(q))$.}
\begin{proof}
  First, let us observe that, by direct induction, given two requests $q$ and $q'$ and a policy $p$, if for any $(n, v) \in \Att(p)$, if $\sem{(n, v)}(q) = \sem{(n, v)}(q')$, then $\sem{p}(q) = \sem{p}(q')$. 

  We now proceed in two steps: we first show that given any $(n, v)$ in $\Att(p)$ if $\sem{(n, v)}(q) = d$, then 
  $\sem{(n, v)}(\nf(q)) = d$, and then we show that the converse holds, which allows us to conclude. 

  Given any $(n, v) \in \Att(p)$, three cases are possible for $\sem{(n, v)}(q)$: 
  \begin{compactitem}
  \item $\sem{(n, v)}(q) = \match$, and in this case $(n, v) \in q$, meaning that $(n, v) \in \nf(q)$, and therefore that $\sem{(n, v)}(\nf(q)) = \match$. 
  \item $\sem{(n, v)}(q) = \nomatch$, and in this case, we first have $(n, v) \not \in q$, implying that $(n, v) \not \in \nf(q)$. Furthermore, there exists $v'$ such that $(n, v') \in q$, and thus either $(n, v') \in \Att(p)$, in which case $(n, v')$ also belongs to $\nf(q)$ and we can conclude that $\sem{(n, v)}(\nf(q)) = \nomatch$, or $(n, v') \not \in \Att(p)$, in which case $(n, \fv(n)) \in \nf(q)$. Since $fv_n \neq v$, we can conclude that $\sem{(n, v)}(\nf(q)) = \nomatch$. 
  \item $\sem{(n, v)}(q) = \noreqmatch$, and in this case, there does not exists any $v'$ such that $(n, v')$ belongs to $q$, meaning that there does exist any $v'$ such that $(n, v')$ belongs to $\nf(q)$, and so we can conclude
    that $\sem{(n, v)}(\nf(q)) = \noreqmatch$. 
  \end{compactitem}

\medskip
\noindent
  Similarly, given any $(n, v) \in \Att(p)$, three cases are possible for $\sem{(n, v)}(\nf(q))$: 
  \begin{compactitem}
  \item $\sem{(n, v)}(\nf(q)) = \match$, and in this case $(n, v) \in \nf(q)$, meaning that $(n, v) \in q$ (since $f \neq \fv(n)$ and therefore that $\sem{(n, v)}(q) = \match$. 
  \item $\sem{(n, v)}(\nf(q)) = \nomatch$, and in this case, we first have $(n, v) \not \in \nf(q)$, implying that $(n, v) \not \in q$. Furthermore, there exists $v'$ such that $(n, v') \in \nf(q)$, and thus either $(n, v') \in q$, in which case we can conclude that $\sem{(n, v)}(q) = \nomatch$, or $v' = \fv(n)$, in which case, there exists $v''$ such 
    that $(n, v'') \in q$, and we can also conclude that $\sem{(n, v)}(q) = \nomatch$. 
  \item $\sem{(n, v)}(q) = \noreqmatch$, and in this case, there does not exists any $v'$ such that $(n, v')$ belongs to $\nf(q)$, meaning that there does exist any $v'$ such that $(n, v')$ belongs to $q$, and so we can conclude
    that $\sem{(n, v)}(q) = \noreqmatch$. 
  \end{compactitem}
\end{proof}

\noindent
\textbf{Proposition~\ref{thm:inclusion}} \textit{
Given any requests $q$ and $q'$ and any policy $p$, if $q' \subseteq q$, then $\nf_p(q') \subseteq \nf_p(q)$. }
\begin{proof}
Let $q$ and $q'$ be two requests such that $q' \subseteq q$, let $(n, v) \in \nf(q')$ and let us show that $(n, v) \in \nf(q)$. Indeed, two cases are possible: 
\begin{compactitem}
\item either $(n, v) \in q'$, which means that $(n, v) \in \Att(p)$. Since $q' \subseteq q$, we know that $(n, v) \in q$, and we can therefore conclude that $(n, v) \in \nf(q)$; 
\item or $(n, v) \not \in q'$, which means that $v = \fv(n)$. In this case, there exists $v'$ such that $(n, v') \not \in \Att(p)$  and $(n, v') \in q'$. Since $q' \subseteq q$, we also have $(n, v') \in q$, and since $(n, v') \not \in \Att(p)$, we can conclude that $(n, \fv(n)) \in q$. 
\end{compactitem}
\end{proof}

\noindent
\textbf{Proposition~\ref{thm:restriction}} \textit{
A policy $p$ is resistant if, and only if, given $q$ and $q'$ in $\NF_p(\Q)$, 
if $q' \subseteq q$ and if $\sem{p}(q') = \set{\allow}$, then $\sem{p}(q) = \set{\allow}$. }
\begin{proof}
Since $\NF_p(\Q) \subseteq \Q$, the implication is immediate. We show that it is enough to restrict our attention to $\NF_p(\Q)$ by contraposition: if $p$ is not resistant, then there exists $q$ and $q'$ such that $q' \subseteq q$, 
$\sem{p}(q') = \set{\allow}$ and $\sem{p}(q) \neq \set{\allow}$. From Proposition~\ref{thm:inclusion}, we have 
that $\nf(q') \subseteq \nf(q)$, and from Proposition~\ref{thm:nf}, we have that 
$\sem{p}(\nf(q')) = \set{\allow}$ and $\sem{p}(\nf(q)) \neq \set{\allow}$, thus allowing us to conclude. 
\end{proof}

\noindent
\textbf{Proposition~\ref{thm:step}}\textit{
  A policy $p$ is resistant if, and only if, for any request $q$ such that $\sem{p}(q) \neq \set{\allow}$, if $\sem{p}(q \setminus \set{(n, v)}) \neq \set{\allow}$, for any attribute $n$ and any value $v$. }
\begin{proof}
  Since $(q \setminus \set{(n, v)}) \subseteq q$, the implication is trivial. Now, consider a policy $p$ such that for any $q$ such that $\sem{p}(q) \neq \set{\allow}$, we have $\sem{p}(q \setminus \set{(n, v)}) \neq \set{\allow}$, for any attribute $n$ and any value $v$, and let us show that $p$ is resistant. 

Let $q, q'$ be two requests such that $q' \subseteq q$ and $\sem{p}(q) \neq \set{\allow}$, and let us show that 
$\sem{p}(q') \neq \set{\allow}$. In order to do so, we prove by induction that for any $k$, for any $(n_i, v_i)$, with $0 \leqslant i \leqslant k$, we have
$\sem{p}(q \setminus \set{(n_1, v_1), \dots, (n_k, v_k)}) \neq \set{\allow}$, which will allow us to conclude, since $q' \subseteq q$. 

When $k = 0$, we have $q \setminus \emptyset = q$, and we can conclude by the assumption that $\sem{p}(q) \neq \set{\allow}$. Let us now write $q'' = q \setminus \set{(n_1, v_1), \dots, (n_k, v_k)}$ and 
assume that $\sem{p}(q'') \neq \set{\allow}$; we can show that $\sem{p}(q'' \setminus \set{(n_{k+1}, v_{k+1})}) \neq \set{\allow}$ by assumption over $p$, thus allowing us to conclude our induction proof, and therefore that $p$ is resistant. 
\end{proof}

\section{External files}
\label{app:files}
The repository \url{http://www.morisset.eu/atrap/} lists the file {\bf header.maude}, which contains all the maude definitions used in \ATRAP; {\bf logic.thy}, which contains the encoding of the 3-valued logic in Isabelle; {\bf ptacl.thy}, which contains the encoding of PTaCL in Isabelle; {\bf atrap.thy}, which contains the encoding of the proof techniques described in Section~\ref{sec:resistance} and {\bf p2.thy}, which is the proof generated in Section~\ref{sec:certification}. 
\end{document}